# Phase transition in the maximum clique problem: the case of Erdős-Rényi graphs

Kazuhito Shida

*Tohoku University Biomedical Engineering Research Organization (TUBERO)*

*980-8575, Sendai, Japan*

:

Abstract

A phase transition, like the one already found on Boolean satisfiability problem by Kirkpatrick and Selman, is found on another NP-complete problem: maximum clique problem. The most general class of random graph, Erdős-Rényi graph, is used as the source of random problem instances. The transition exhibits finite size scaling and concentration of the "difficult" graph instances, which need particularly large CPU time to solve, around the transition point.

1. Introduction

  Statistical physics deals with collective phenomena manifested in a large population of interacting elements in a universal manner. When a complicated problem is solved by a computer, the algorithm and data structure for the problem must be represented in the computer by a large population of bits, which are "interacting" because their values are constrained in a manner specific to the problem. Thus, it may be possible to create a "statistical physics of computational problem solving" to analyze the ensemble behavior of given type of computational problem.

  Such a link between computational processes and statistical physics was first investigated by Fu and Anderson back in 1985[1], where they discuss the possible links between the energy optimization of the rough potential landscapes of spin-grass and a class of computational problem which demands a particularly large number of computational steps to solve, called nondeterministic polynomial (NP)-complete problems. Despite the extensive efforts of many researchers, no polynomial-time algorithms are known for any of NP-complete problems, and NP-complete problems are usually regarded as "intractable".

  Recently, a series of new examples of "statistical physics of computational problem solving" have been realized: it has been reported [2-4] that an NP-complete problem called *k*-SAT exhibits a "phase transition". In a *k*-SAT problem, the satisfiability of a conjunctive normal form (CNF) with *k* Boolean variables per clause (e.g. $(\bar{x} \vee y \vee z) \wedge (x \vee \bar{y} \vee \bar{z})$ for *k*=3) has to be determined. This is the first problem to be proven as NP-complete for *k*>2[5]. When numerous randomly generated *k*-SAT problems are solved, however, the truly difficult problems associated with large numbers of computational steps are concentrated near a critical value of the parameter $\alpha = M/N$, where *M* and *N* are the number of clauses and variables, respectively. Moreover, the critical point is accompanied by a sharp increase of overall probability that the expression cannot be satisfied: This is a transition of "difficulty" of computational problems. The transition becomes increasingly sharp for larger values of *N*, and surprisingly, a finite size scaling can be established on k-SAT problem regarding *N* as the "system-size".

Although there is no rigorous mathematical proof of the equivalence between phase transitions in thermodynamic systems and combinatorial problems, the analogy between them is evident. The relationship between the number of steps required for the algorithm and the correlation in thermodynamic systems (in other words, smoothness in the energy landscape) is intuitively understandable. Such a phase transition have been established for couples of [6][7] NP-complete problems and suggesting a very interesting possibility for efficient solutions of large combinatorial problems.

This report presents a nontrivial extension of these results to another NP-complete problem: the maximum clique problem. The maximum clique problem frequently appears in practical applications in knowledge engineering, complexity science and bioinformatics. Established features of this important problem include a sharp transition in clique occurrence probability as a function of clique size, *q*, finite-size scaling, and a rather sharp peak in computational complexity. To establish the most universal characteristic of maximal clique discovery for these broad subjects, the most general class of random graphs, the Erdős-Rényi (ER) graph, is used as the ensemble. It is also proposed later that the finite-size scaling can be used in a totally new way in some of those applications.

2. Method

In a maximum clique problem on a graph $G = (V, E)$, one has to find the largest subset $S \subset V$ such that for all $v_x, v_y \in S$, $(v_x, v_y) \in E$: the largest subgraph of the given input in which all vertices are mutually connected. Such a subset is called the maximum clique, and the number of vertices in the maximum clique is called the *clique number*. This maximum clique problem (sometime called max-clique), is included in the first 21 problems proven by Karp [8] to be NP-complete by means of the reduction to *k*-SAT.

In ER graph, edges are placed on $|E|$ positions randomly selected from $|V|(|V|-1)/2$ possible positions: the edges have no correlation among them (There is no preference of edge location depending locations of other edges, as observed in the small world networks). Because there is no correlation modeled by the clustering coefficient [9], the random distribution of the edges is completely specified by the edge density, $p$, defined as $2|E|/|V|(|V|-1)$.

Phase transition of maximum clique problem is nontrivial and it must be discussed independently from the phase transition of *k*-SAT. Although the proof of NP-completeness of the maximum clique problem was based on a mapping between random graphs and random CNFs, it seems difficult to utilize the mapping to predict the ensemble behavior of maximum clique problem due to several restrictions on the mapping. For instance, it is difficult to generate graphs with edge density greater than 1/2, because every edge is a representation of the "simultaneous satisfiability" between pairs of variables in k-SAT problems, as in $x \leftrightarrow y$ or $\bar{x} \leftrightarrow y$, but not in $\bar{x} \leftrightarrow x$. Since this is a severe restriction on statistical analysis on ER graphs, all investigation had to be carried out in an empirical manner.

3. Result

Random graph ensembles were generated for a number of combinations of $p$ and $|V|$ given in Table 1, according to a Mersenne twister pseudo random number generator[10]. Typically, $10^4$ independent ER graphs were prepared for a given $p$ and $|V|$ as an ensemble. ER graphs are generated by a simple random edge generation, which means that graph isomorphism is not considered in the preparation of the graph ensemble. This is not because the exclusion of isomorphism is difficult (although no polynomial-time algorithm is known for identification of isomorphism), but mainly because CNFs with apparent identity were not excluded in the SAT transition paper.

The maximum clique problem is solved for each graph in the ensemble by the CLIQUER[11] algorithm. CLIQUER is a very efficient clique solver based on an advanced prediction of the clique number. The CPU time required for CLIQUER to obtain a maximum clique was used as the index of computational complexity. All clique discovery was performed on four of Intel Core2Duo computers with identical specifications running on LINUX. The largest graph solved was constituted by 702 vertices and $\sim 7.42 \times 10^4$ edges, and consecutive CPU time spend is about two months.

A crucial element of the investigation of phase transitions in computation is of course the selection of the order parameter. Here, the order parameter of choice is the fraction of graph instances devoid of cliques with given size, denoted as $UNC(q)$, which stands for "UN-Clique" . This parameter was chosen over the other definition, $CLQ(q) = 1 - UNC(q)$, such that the transition plot can be directly compared to the plots in previous works (such as Fig. 3 of Kirkpatrick and Selman).

The transition point $q'_c$ is defined on the normalized clique number, $q' = q/\bar{q}(p,|V|)$. The normalization factor $\bar{q}(p,|V|)$ is the clique at which $UNC(q)$ exceeds some threshold. Bollobás has already proposed [12] that the mean clique size scales as

$$2\log_{1/p}|V| - 2\log_{1/p}\log_{1/p}|V| + 2\log_{1/p}(e/2) + 1 + O(1), \qquad (1)$$

for random graphs. As shown later, this is a good approximation only when $p$ is small. In addition, "mean clique size" may not be a good normalization factor if probabilistic distribution of clique size is skewed.

The value of $\bar{q}(p,|V|)$ which makes $CLQ(q) = UNC(q) = 0.5$ can be given approximately by solving the following equation:

$$P^{(|V|(|V|-1)-\bar{q}(\bar{q}-1))/2}_{(|E|-\bar{q}(\bar{q}-1))/2} P^{|V|}_{\bar{q}} / P^{|V|(|V|-1)/2}_{|E|/2} = 0.5. \qquad (2)$$

The denominator in the equation is the number of all possible graphs for given $|V|$ and $|E|$. The numerator in the equation is the number of placing $|E|$ edges in $|V|(|V|-1)/2$ places when $\bar{q}(\bar{q}-1)/2$ edges are reserved for a $k$-clique: The number of all possible graphs for given $|V|$ and $|E|$, whose clique numbers are at least $k$. Although the clique size is an integer, a real number solution of $\bar{q}(p,|V|)$ can be numerically obtained by replacing factorials $P^n_m$ by $\Gamma(n-1)/\Gamma(n-m-1)\Gamma(m-1)$ and scanning the value of $\bar{q}$ with an appropriate interval. All data introduced in this paper other than Figure 1 is rescaled by $\bar{q}(p,|V|)$ obtained by this method (interval = 0.1).

Figure 1 shows the relationship between $|V|$ and the mean size of maximum clique. The dashed lines are the values given by the approximation formula by Bollobás, which is clearly underestimating the clique size. It is justified by this plot to use the definition of normalized clique size, $q'$, based on Eq. 2, shown by the solid lines. Note that $|V|$ is large enough for current objective: the clique sizes observed in many case are typically greater than 10 for p=0.5, 0.7 and 0.9, which helps to treat the clique number distributions as continuous ones.

Figure 2 shows the relationship between the normalized clique size versus the value of $UNC(q)$ for various combinations of $p$ and $q'$. The transition to the non-Clique phase is increasingly sharp for larger values of $|V|$, which strongly suggests that this is a SAT-like phase transition with $|V|$ being the system size. Note the sparseness of the data points in the figure: the distribution of the clique number of the ER graph tends to be very narrow, in particular for small values of $p$. This render it extremely difficult to evaluate the critical point and critical exponent, particularly when *p=0.3* and *0.5*.

For the maximum-clique problem, the standard form of the finite-size scaling formula[13] can be adopted in the form of

$$q'_{FSSed} = |V|^{1/\nu} (q' - q'_c) / q'_c. \qquad (3)$$

The values of $q'_c$ and $\nu$ are adjusted for various values of $p$, such that the overlap between rescaled $UNC(q_{FSSed})$ plots is maximized for *p=0.9*. In Figure 3, the value of $UNC(q_{FSSed})$ is plotted against the rescaled $q_{FSSed}$, with $q'_c$=0.96 and $\nu$=2.0. Although the data points in the figure are rather sparse (due to the sharp increase of $UNC(q)$) for small values of *p*, the plots clearly show successful finite-size scaling for p=0.5 and greater. In particular, the finite size scaling in the case of $p = 0.9$ is doubtless. For *p=0.3*, it is either $q'_c$=0.96, $\nu$=2.0 may be a wrong parameter setting or FSS is impossible altogether. The value range of the system's size parameters is narrower than that found in the FSS of, for example, percolation transition, but it is comparable to that reported for the FSS of *k*-SAT.

The similarity between k-SAT and maximum clique problems regarding their phase transition is further supported by a plot of the normalized clique size versus mean CPU-time to solve the maximum-clique problem. In Figure 4, "difficult" graphs are clearly concentrated at the border between the two phases when graphs are solved by

CLIQUER. This trend cannot be addressed to the characteristics of CLIQUER algorithm, because the same trend is observed in additional tests performed by a slower but more standard DFMAX[14] algorithm. It seems, as observed in the k-SAT and TSP problems, all really difficult maximum clique problems (that is, all *graphs* really difficult in terms of maximum clique finding on them) are actually limited in a narrow region in the problem space. This observation has a certain impact on any maximum clique related problems on graphs with large number of edges.

Kirkpatrick and Selman[2] noted in their paper that "Yet there are other NP-complete problems (for example, the traveling salesman and max-clique) that lack a clear phase boundary at which hard problems cluster"; however, as shown in this paper, both a phase boundary and hard problem concentration near the boundary can be discovered for max-clique.

It seems worthwhile to consider what might be the reason of the error in their observation on the max-clique transition. One possible explanation is that they were looking for the transition in the wrong region of $p$. The maximum clique transition is very difficult to characterize when $p$ is too small, and intractably time-consuming when $p$ is too large. A breakdown of simple approximation of the transition point for relatively large $p$ values might have been another reason. The most interesting cause of the error to consider is the algorithm they were using to solve the max-clique problem. The max-clique algorithm for this type of work must be highly efficient. Because not only a large problem ensembles have to be solved, but also their execution time must correctly reflect the inherent level of difficulty of the given problem (otherwise, the cluster of hard problems becomes invisible).

4. Discussion

In this report, a new phase transition was identified in the maximum-clique discovery problem for the most basic class of random graphs. In many respects, the characteristics of the phase transition are close to those of transitions already established on other NP-complete problems, namely k-SAT. The most notable element in the closeness between maximum clique and k-SAT transitions was the finite-size scaling, which may link the current result further to more "physical" phase transitions like the Ising model and percolation models.

Of course, this is not the first phase transition found on random graphs. Erdős and Rényi established a transition in the connectivity[15] of random graphs. Derényi, Palla and Vicsek found a percolation-like transition[16] in the connectivity between small cliques in random graphs. However, this result is the first one linking the rareness of cliques of the given size and the computational cost of finding them. This may have many implications for various fields of science dealing with random graphs and networks.

Whenever complex network of relationship among many objects is analyzed, the ER graph works as the most basic null-hypothesis of their connection pattern because it does not assume any correlation or order between edges. Maximum clique can be a useful idea in network science because it is the best model of objects in a particularly strong relationship in a complex network: a "pattern" in a random graph. Now, a rule with high universality is revealed in the probabilistic distribution of the clique number, in a form closely related to the computational cost to find them. Such a rule will be quite useful for improving the efficiency of large-scale network analyses, which can be very computationally intensive. Also, this rule is useful in evaluating the statistical significance of the connection patterns found as cliques.

In particular, many applications of the current result are anticipated in the field of bioinformatics, where complex relationship among numerous biological objects (genes, proteins, ligands, and so on) must be scrutinized. For example, maximum clique is used as the model of a "hidden pattern" in sequence motif finding[17] and in-silico pharmacophore screening[18]. In bioinformatics, whether a biological pattern (e.g. similar sequences found near corresponding human and mouse genes, called sequence motif) found in biological data is worth further analysis (e.g. molecular biological experiments like mutagenesis) or not should be determined by the statistical significance of the pattern, indexed by P-value, which is the probability that the patterns with at least given strength appears purely by coincidence. The precise evaluation of P-value often requires troublesome numerical integration. However, this can be much simplified by means of finite-size scaling because the scaling rule enables us to translate between clique rareness profiles for different vertex sizes.

Many problems, though, remain to realize these applications. First, we do not know how wide the effective regime of finite-size scaling is in the current problem. And testing

P-value calculations must be performed where patterns or cliques are fairly rare; performing it with enough precision requires a very large-scale computational experiment. The other major problem is the "rigidity" of the current model. The weighted edges and maximum density subgraph (the subgraph with maximized sum of edge weight per node) are believed to be a much better and flexible model of hidden patterns. Constructing a theory similar to that developed in this work on maximum density subgraphs should be possible because it has already been proven that finding dense subgraphs in a weighted graph is still NP-complete[19], if the target subgraph is sufficiently "dense".

Acknowledgements

This work was supported by Special Coordination Funds for Promoting Science and Technology from the Japanese Ministry of Education, Culture, Sports, Science and Technology.

References

1       Y. T. Fu and P. W. Anderson, Journal of Physics A-Mathematical and General **19**, 1605 (1986).
2       S. Kirkpatrick and B. Selman, Science **264**, 1297 (1994).
3       R. Monasson, R. Zecchina, S. Kirkpatrick, et al., Nature **400**, 133 (1999).
4       D. Achlioptas, A. Naor, and Y. Peres, Nature **435**, 759 (2005).
5       S. Cook, in *Proceedings of the third annual ACM symposium on Theory of computing* (ACM, 1971), p. 151–158.
6       I. P. Gent and T. Walsh, Artificial Intelligence **88**, 349 (1996).
7       A. K. Hartmann, W. Barthel, and M. Weigt, Computer Physics Communications **169**, 234 (2005).
8       R. Karp, in *Complexity of Computer Computations*, edited by R. Miller and J. Thatcher (Plenum Press, 1972), p. 85.
9       D. J. Watts and S. H. Strogatz, Nature **393**, 440 (1998).
10      M. Matsumoto and T. Nishimura, ACM Trans on Modeling and Computer Simulations **8**, 3 (1998).
11      P. Ostergard, Discrete applied mathematics **120**, 197 (2002).
12      B. Bollobás, *Random graphs* (Cambridge university press, 2001).
13      M. Fisher and M. Barber, Physical review letters **28**, 1516 (1972).

14  R. Carraghan and P. M. Pardalos, Operations Research Letters **9**, 375 (1990).
15  P. Erdős and A. Rényi, Acta Mathematica Academiae Scientiarum Hungaricae **17**, 359 (1966).
16  I. Derényi, G. Palla, and T. Vicsek, Physical Review Letters **94** (2005).
17  P. A. Pevzner and S. H. Sze, Proc Int Conf Intell Syst Mol Biol **8**, 269 (2000).
18  T. J. Ewing, S. Makino, A. G. Skillman, et al., J Comput Aided Mol Des **15**, 411 (2001).
19  Y. Asahiro, R. Hassin, and K. Iwama, Discrete Applied Mathematics **121**, 15 (2002).

Table and figures

| $p$ | $|V|$ | Ensemble size |
|---|---|---|
| 0.30 | 64, 128, 256, 320, 384, 448, 512, 576, 640, 704 | $10^4$ |
| 0.50 | 64, 128, 160, 192, 224, 256, 288, 320, 352 | $10^4$ |
| 0.70 | 64, 96, 112, 128, 144, 160, 176, 192 | $10^4$ |
| 0.90 | 64, 96, 112, 128, 144 | $10^4$ |

Table 1. Combinations of $p$ and $|V|$ investigated in this work.

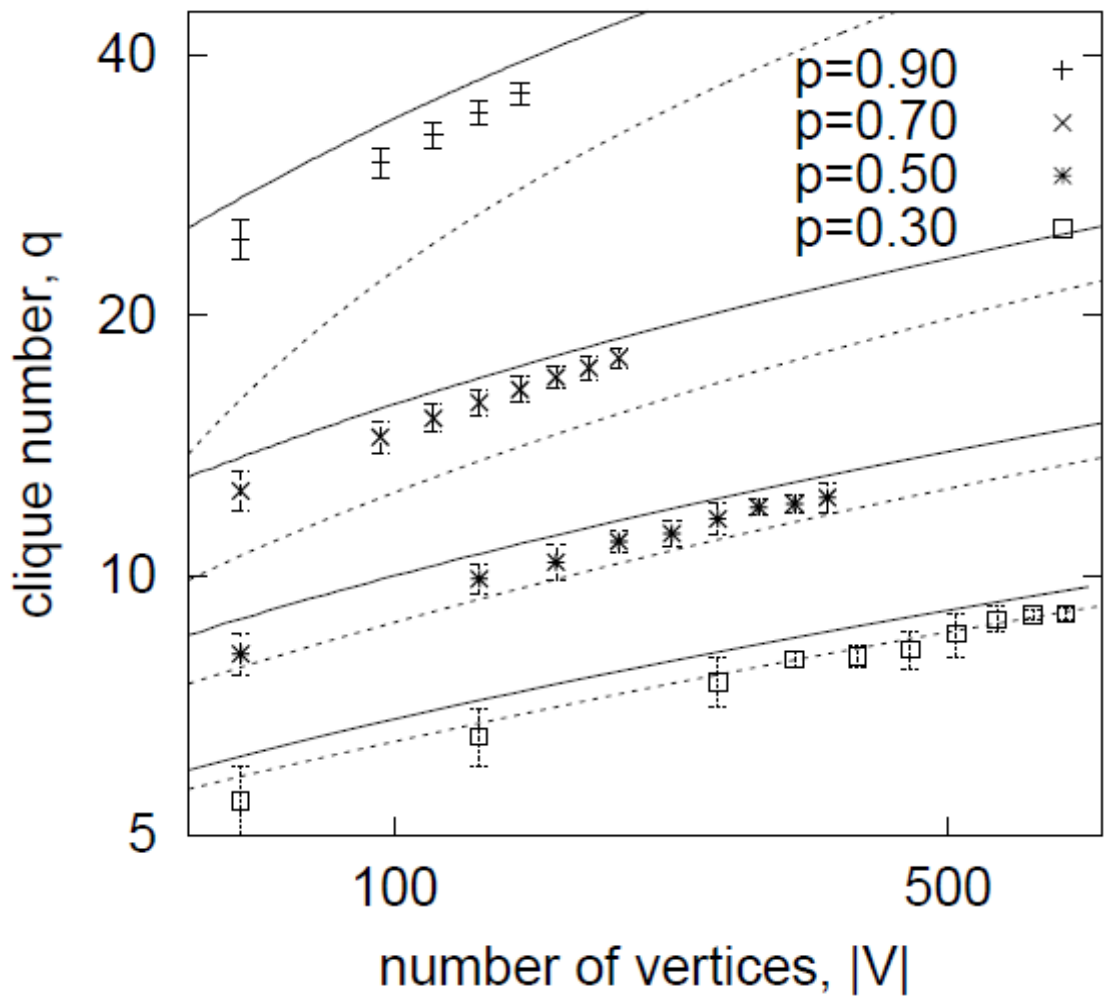

Figure 1 : The mean and standard deviation of maximum clique size obtained for various |V| and p. The dotted lines are the Bollobás approximation and the solid lines are the approximation given by Eq. 2 for p=0.9, 0.7, 0.5, and 0.3, from above, respectively.

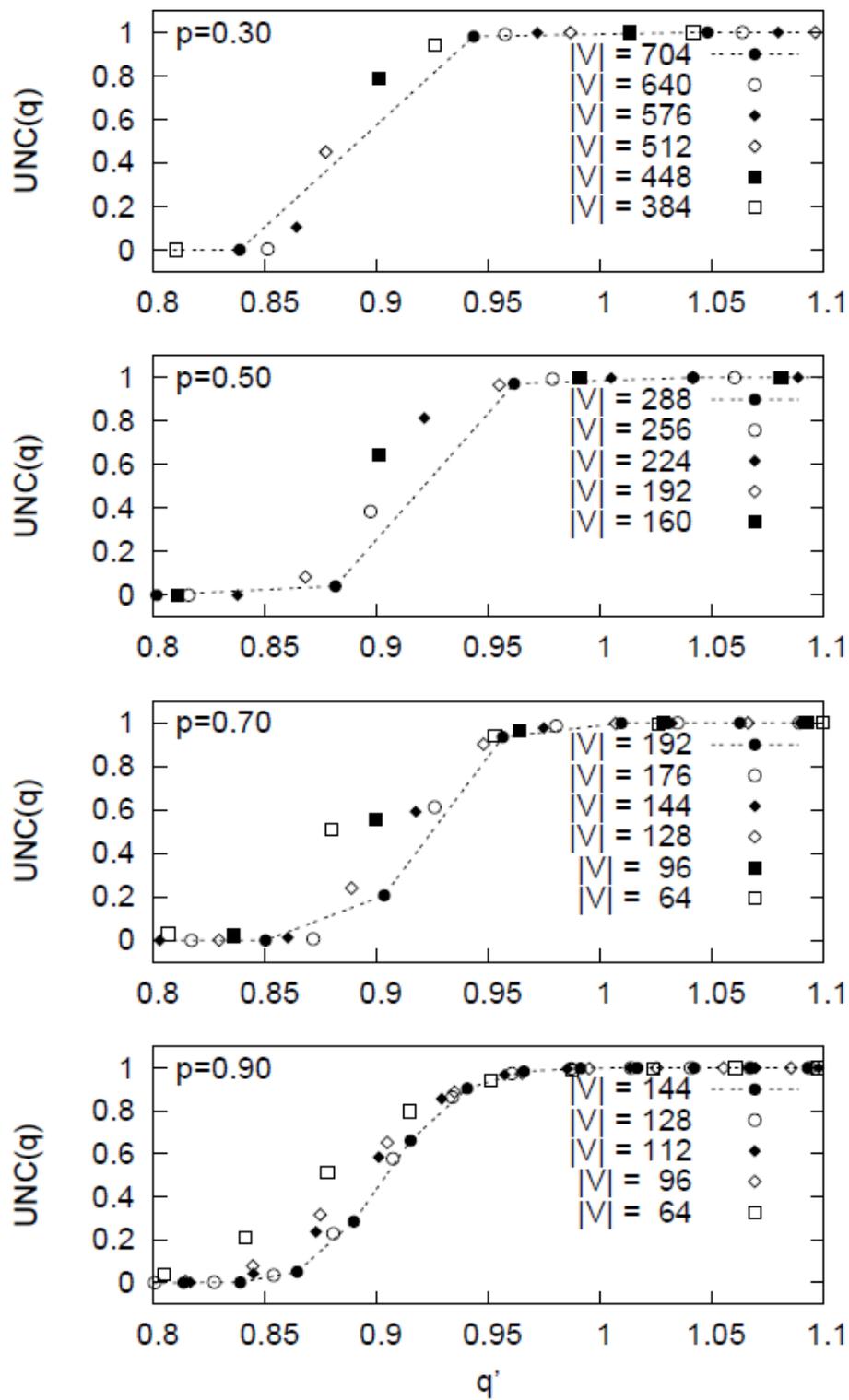

Figure 2 : Normalized maximum clique size versus the $UNC(k)$ in ER graphs. Large |V| is relevant to shaper transitions.

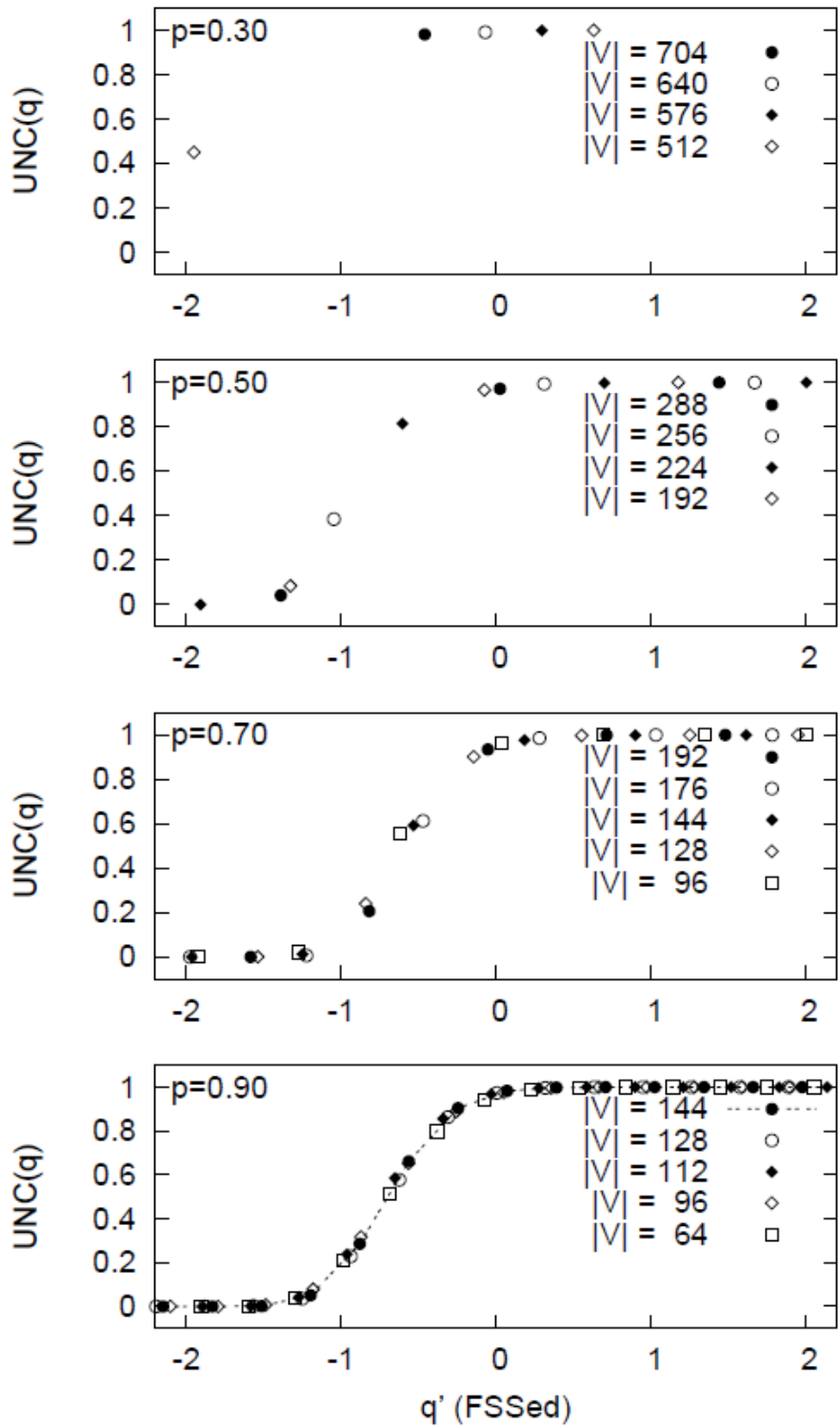

Figure 3 : Finite-size scaled maximum clique size versus the $UNC(k)$ in ER graphs. A phase transition and nice finite-size scaling is observed in max clique problem in ER-graph

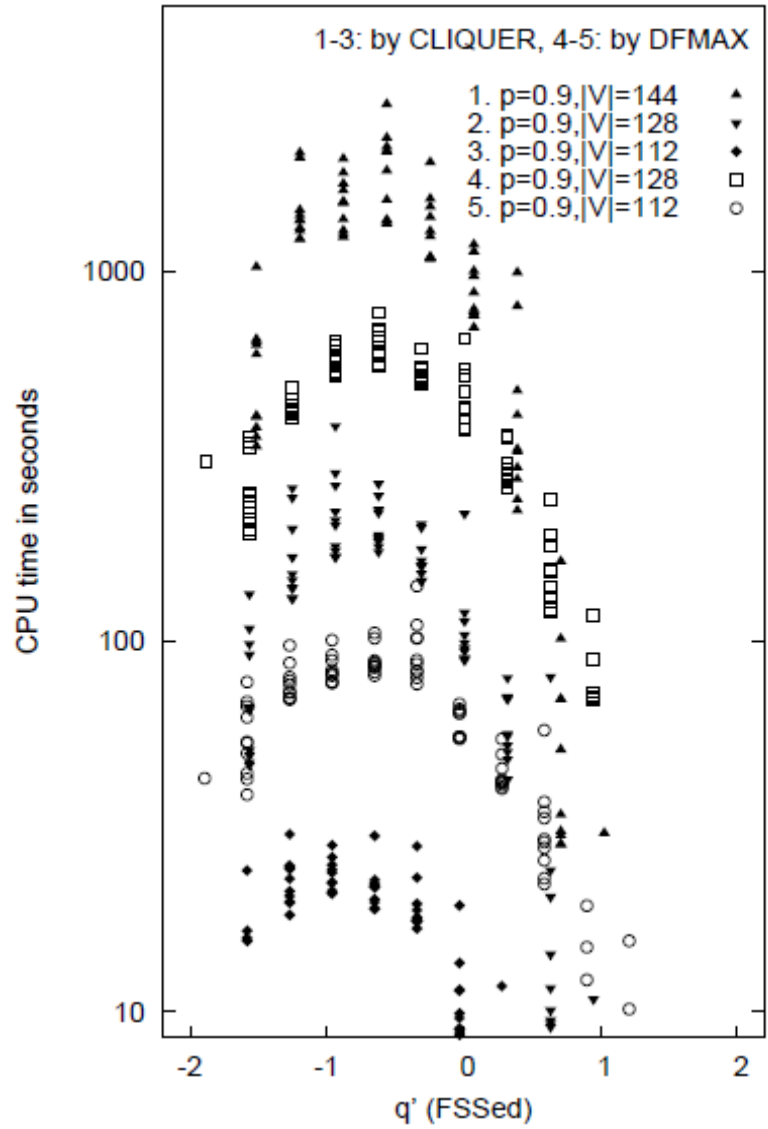

Figure 4 : The 10 largest CPU-time to solve the maximum-clique problem for various $p$ and $q$ is shown with x-axis identical to Figure 3 and y-axis in seconds (log manner). Note that Graphs with large computational complexity tend to be concentrated near the phase boundary.